# New Ternary Superconducting Compound LaRu$_2$As$_2$: Physical Properties from DFT Calculations

M. A. Hadi[1*], M. S. Ali[2], S. H. Naqib[1] and A. K. M. A. Islam[1,3]

[1]Department of Physics, University of Rajshahi, Rajshahi-6205, Bangladesh
[2]Department of Physics, Pabna University of Science and Technology, Pabna-6600, Bangladesh
[3]International Islamic University Chittagong, Chittagong-4203, Bangladesh

**Abstract**

In this paper, we have presented the density functional theory (DFT) based calculations performed within the first-principles pseudopotential method to investigate the physical properties of the newly discovered superconductor LaRu$_2$As$_2$ for the first time. The optimized structural parameters are in good agreement with the experimental results. The calculated independent elastic constants ensure the mechanical stability of the compound. The calculated Cauchy pressure, Pugh's ratio as well as Poisson's ratio indicate that LaRu$_2$As$_2$ should behave as a ductile material. Due to low Debye temperature, LaRu$_2$As$_2$ may be used as a thermal barrier coating (TBC) material. The new compound should exhibit metallic nature as its valence bands overlap considerably with the conduction bands. LaRu$_2$As$_2$ is expected to be a soft material and easily machineable because of its low hardness value of 6.8 GPa. The multi-band nature is observed in calculated Fermi surface. A highly anisotropic combination of ionic, covalent, and metallic interactions is expected in accordance with charge density calculations.

## 1. Introduction

A great variety of compounds with RT$_2$X$_2$ stoichiometry take on the body-centered tetragonal ThCr$_2$Si$_2$ type structure (space group $I4/mmm$, No. 139) [1]. In this stoichiometry, R is symbolized for a rare earth, alkaline earth or alkali element, T refers a transition metal and X represents p-metal atoms, namely B, P, Si, As or Ge. This large family includes more than 700 members of 122-type intermetallics (so-called 122 phases) and exhibits an outstanding collection of physical and chemical properties [2]. The structure of RT$_2$X$_2$ stoichiometry offers an innate multilayer system in which the planes of R ions are taken apart from T metallic layers by X atomic sheets, comprising of R–X–T–X–R stacking of individual basal planes along the c-axis. The ThCr$_2$Si$_2$ type structure exhibits a strong uniaxial anisotropy with keeping c-axis as an anisotropy axis. In other words, the layer of edge sharing TX$_4$ tetrahedron that is parallel to the *ab* plane and separated by the basal plane of R atoms characterizes the structure of RT$_2$X$_2$. In this structure, the R atomic sheets separate two adjacent [T$_2$X$_2$] blocks. In sequence, T ions inside [T$_2$X$_2$] blocks build up a square lattice pressed between two X sheets shifted in order that each T is enclosed by a distorted X tetrahedron TX$_4$. A wide range of variation occurs in the interlayer distances between X atoms when either R or T is altered for a particular p-metal atom [2,3].

This group of materials have attracted a lot of attention in recent years because of the discovery of high-temperature superconductivity in iron arsenide AFe$_2$As$_2$ (A = K, Ca, Sr, Ba, etc.) [4,5]. As Ru lies in the same group together with Fe in the periodic table, the superconductivity in the level of iron arsenide is anticipated for Ru-related compounds, and several studies have been devoted to these materials [6-10]. Motivated by this, Guo et al. [11] have recently synthesized the polycrystalline samples of ThCr$_2$Si$_2$-type LaRu$_2$As$_2$ and reported the bulk superconductivity in LaRu$_2$As$_2$. Their temperature dependent resistivity measurement ensures a superconducting transition of LaRu$_2$As$_2$ with an onset $T_c$ of 7.8 K. This measurement also estimates the upper critical field $\mu_0 H_{c2}(0)$ at zero temperature and obtains a value of 1.6 T. Further, DC magnetic susceptibility measurement detects a bulk superconducting Meissner transition at 7.0 K, and the isothermal magnetization measurement signifies that LaRu$_2$As$_2$ is a type-II superconductor. Surprisingly, LaRu$_2$As$_2$ has the highest transition temperature $T_c$ among the iron-free transition metal pnictides with the ThCr$_2$Si$_2$-type crystal structure.

In this paper, the study of LaRu$_2$As$_2$ has been carried out via the ground state electronic structure calculations using the plane-wave pseudopotential approach within the density functional formalism. The aim of the present paper is to investigate the ground state properties including structural, elastic, electronic and bonding characters of the newly discovered 122 phase compound by calculating its total energy at optimized volume. To the best of our knowledge, this is the first *ab initio* calculations completed on this new superconductor $LaRu_2As_2$.

---

[*]Corresponding author:hadipab@gmail.com

The rest of this paper is divided into three sections. In Section 2, a brief description of the method of computation used in this study has been presented. The results obtained for the structural, elastic, electronic, and bonding properties of $LaRu_2As_2$ are analyzed in Section 3. Finally, Section 4 consists of the main conclusions of the present work.

## 2. Method of calculations

The approach used to investigate the physical properties of $LaRu_2As_2$ as implemented in the Cambridge Serial Total Energy Package (CASTEP) code [12] is as follows: a set of Kohn-Sham equations are solved by means of the plane-wave pseudopotential method [13] within the strong electron ensemble Density Functional Theory (DFT) [14]. Exchange and correlation energies are described by a nonlocal correction for LDA in the form of GGA [15]. The wavefunctions are allowed to expand in a plane wave basis set followed by the periodic boundary conditions and Bloch's Theorem [16]. The electron-ion potential is treated by means of first-principles pseudopotentials within Vanderbilt-type ultrasoft formulation [17]. BFGS energy minimization technique [18] best for crystalline materials is used to find, self-consistently, the electronic wavefunctions and consequent charge density. In addition, the density mixing [19,20] scheme is employed. For *k*-points sampling integration over the first Brillouin, the Monkhorst–Pack scheme [21] is used. The parameters are chosen for present calculations are as follows: energy cutoff 500 eV (340 eV for elastic properties calculations), *k*-point mesh 16 × 16 × 7, the difference in total energy per atom within $5 \times 10^{-6}$ eV, maximum ionic Hellmann-Feynman force within 0.01 eV/Å, maximum stress within 0.02 GPa, and maximum ionic displacement within $5 \times 10^{-4}$ Å.

## 3. Physical properties

In this section the investigated physical properties of the 122 phase superconductor $LaRu_2As_2$ are presented and analyzed; among them are the structural and elastic properties, Debye temperature, Band structure and DOS, and then Mulliken population and Vickers hardness and the electron charge density as well as Fermi surface of $LaRu_2As_2$ under zero pressure.

### 3.1 Structural properties

The body centered tetragonal $LaRu_2As_2$ structure shown in Fig. 1 crystallizes with space group *I*4/*mmm* (No. 139). Every atomic species in this crystal occupy only one crystallographic position and only one internal parameter $z_X$ settles the relative position of As inside the unit cell. There are two formula units in each unit cell. The 2a, 4d and 4e atomic positions are inhabited by the La, Ru, and As atoms, respectively. The negatively charged $[Ru_2As_2]$ blocks and positively charged La layers take positions by turns along the *c*-direction and form the crystal structure. The optimized lattice constants and $z_X$ parameter of $LaRu_2As_2$ are given in Table 1 along with those obtained for some other 122 phase compounds. The computed lattice parameters accord well with the measured values [11]. Moreover, the calculated structural parameters are slightly larger ($a \sim 0.23\%$, $c \sim 2.94\%$ and $V \sim 3.41\%$) than the experimental values, which is a general trend inherent to GGA calculations.

**Table 1**. Structural properties of $LaRu_2As_2$ in comparison with those for the related $ThCr_2Si_2$ type compounds.

| Compounds | *a* (Å) | *c* (Å) | *V* (Å$^3$) | *c/a* | $z_X$ | References |
|---|---|---|---|---|---|---|
| $LaRu_2As_2$ | 4.1923 | 10.9014 | 191.59 | 2.6003 | 0.3641 | [This] |
| | 4.1826 | 10.5903 | 185.27 | 2.5320 | -- | [11] |
| $BaRu_2As_2$ | 4.1925 | 12.3136 | 216.44 | 2.9371 | 0.3510 | [22] |
| | 4.1525 | 12.2504 | 211.24 | 2.9501 | 0.3527 | [23] |
| $SrRu_2As_2$ | 4.2068 | 11.2903 | 199.81 | 2.6838 | 0.3591 | [22] |
| | 4.1713 | 11.1845 | 194.61 | 2.6813 | 0.3612 | [23] |

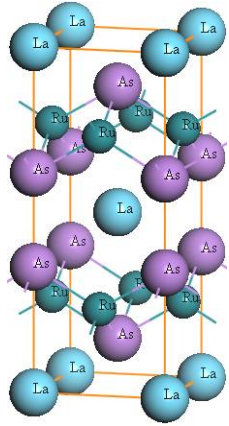

**Fig. 1.** Crystal structure of new 122 phase superconductor $LaRu_2As_2$.

*3.2 Elastic properties*

The mechanical behavior of solids can be described successfully by the elastic constants that are related to materials' response to an applied stress. The finite strain theory assigned in CASTEP code is well established for calculating the elastic constants successfully for a range of materials including metallic systems [24]. In this method, a given uniform deformation $\varepsilon_j$ is applied and then the resulting stress $\sigma_i$ is calculated. By choosing an appropriate functional deformation, elastic constants are then evaluated by solving the linear equation, $\sigma_i = C_{ij}\,\varepsilon_j$.

The well-established Voight-Reuss-Hill (VRH) approximations validated in many metallic and insulating materials [25-27] are applied to calculate the polycrystalline bulk elastic properties namely, bulk modulus $B$ and shear modulus $G$ from calculated $C_{ij}$. Further, the equations, $Y = (9GB)/(3B + G)$ and $v = (3B – 2G)/(6B + 2G)$ are employed to determine the Young's modulus $Y$ and Poisson's ratio $v$, respectively.

The values obtained for single crystal elastic constants ($C_{11}$, $C_{12}$, $C_{13}$, $C_{33}$, $C_{44}$ and $C_{66}$) and bulk elastic properties ($B$, $G$, $Y$, $B/G$ and $v$) of tetragonal crystal system $LaRu_2As_2$ are listed in Table 2 along with the results found in literature for La-based isostructural compounds [28]. The mechanical stability conditions for tetragonal crystals [29] are as follows: $C_{11} > 0$, $C_{33} > 0$, $C_{44} > 0$, $C_{66} > 0$, $C_{11} – C_{12} > 0$, $C_{11} + C_{33} – 2\,C_{13} > 0$, $2(C_{11} + C_{12}) + C_{33} + 4\,C_{13} > 0$. Satisfying these criteria the newly synthesized 122 phase appears as mechanically stable compound. The elastic constants $C_{11}$ and $C_{33}$ explain the linear compression resistance along the a- and c-directions, respectively. It is evident that the obtained elastic constants $C_{11}$ and $C_{33}$ are very large in comparison to other elastic constants, implying that the $LaRu_2As_2$ crystal should be so much incompressible under uniaxial stress along both a- and c-directions. Again, the elastic constant $C_{11}$ is much larger than $C_{33}$, meaning that the incompressibility along the a-direction should be much higher than that along c-direction. In fact, the bonds aligned to the a-axis contribute a dominating effect on $C_{11}$ making it much larger than $C_{33}$. Since $C_{11}+C_{12} > C_{33}$, the bonding in the (001) plane should exhibit more elastic rigidity than that along the c-axis and the elastic tensile modulus should be higher on the (001) plane than that along the c-axis. The elastic constant $C_{44}$ indicates the ability to resist the shear distortion in (100) plane, whereas the elastic constant $C_{66}$ correlates to the resistance to shear in the <110> direction [30]. Since $C_{66} > C_{44}$, the new compound should be more competent to resist the shear distortion in the <110> direction than in the (100) plane. The shear anisotropy factor $A$, defined as $A = 2C_{66}/(C_{11} – C_{12})$ is 1.24, signifying that the shear elastic properties of the (001) plane in $LaRu_2As_2$ depend on the shear directions significantly.

The concept of failure mode i.e., ductile or brittle nature, of materials is important in study of mechanical behaviors of solids. A material can be classified as either brittle or ductile for most practical situations. To quantify the failure state of solids, Cauchy pressure, and Pugh and Poisson's ratios are used as powerful tools. A brittle material changes its volume easily and a ductile material can be easily distorted under the action of external loads. The Cauchy pressure [31] defined as ($C_{12} – C_{44}$) identifies a solid as either brittle or ductile one. When Cauchy pressure is positive (negative), the material is prone to be ductile (brittle). In accordance with the Pugh's ratio [32], the high (low) $B/G$ value signifies the ductile (brittle) nature of the materials. The threshold value of $B/G$ is found to be 1.75, which separates the ductile materials from brittle ones. Frantsevich et al. [33] separated the ductile materials from brittle materials in terms of Poisson's ratio. This rule proposes $v \sim 0.26$ as the border line that separates the brittle and ductile materials. A material having the Poisson's ratio greater than 0.26 will be ductile, otherwise the material will be brittle.

The Cauchy pressure ($C_{12} - C_{44}$) of LaRu$_2$As$_2$ is positive, its Pugh's ratio $B/G$ is 1.79 which is greater than 1.75 and its Poisson's ratio $v$ is 0.265 which is greater than 0.26. As a result, the compound LaRu$_2$As$_2$ should behave as a ductile material. It should be mentioned that there are some papers [34–37] in which $B/G = 2.0$ ($G/B = 0.5$) and $v \sim 0.33$ are used as critical values for separating the brittle materials from ductile ones. On the other hand, many authors [38–44] used the current values presented in this paper to judge brittle/ductile nature of the materials. In fact, Vaitheeswaran et al. [39] prefer $B/G = 1.75$ as threshold value for defining the brittle/ductile nature of the materials and showed that Pugh's critical value corresponds to $v = 0.26$ as the border line between the ductile and brittle materials. These criteria are frequently used to identify the brittle material from ductile ones.

**Table 2.** Single crystal elastic constants $C_{ij}$, polycrystalline bulk modulus $B$, shear modulus $G$, and Young's modulus $Y$, in GPa, Pugh's ratio $G/B$, and Poisson's ratio $v$ of LaRu$_2$As$_2$.

| Compounds | $C_{11}$ | $C_{12}$ | $C_{13}$ | $C_{33}$ | $C_{44}$ | $C_{66}$ | $B$ | $G$ | $Y$ | $B/G$ | $v$ | $A$ | Ref. |
|---|---|---|---|---|---|---|---|---|---|---|---|---|---|
| LaRu$_2$As$_2$ | 244 | 86 | 82 | 113 | 62 | 98 | 113 | 63 | 159 | 1.79 | 0.265 | 1.24 | [This] |
| LaNi$_2$Ge$_2$ | 152 | 87 | 40 | 126 | 87 | 90 | 62 | 74 | 159 | 0.84 | 0.073 | 2.77 | [28] |
| LaNi$_2$P$_2$ | 202 | 116 | 15 | 102 | 116 | 97 | 62 | 93 | 186 | 0.67 | 0.002 | 2.26 | [28] |

For describing the elastic behavior of solids, bulk and shear moduli are two important parameters. Fracture and plastic deformation are two essential issues related to solid materials. Bulk modulus $B$ measures the resistance to fracture and shear modulus $G$ evaluates the resistance to plastic deformation of polycrystalline materials [45]. From dislocation theory [46] it is known that a weak relationship is observed between bulk modulus and hardness. On the contrary, a good relationship exists between hardness and shear modulus [47]. Certainly, the hardness is less susceptible to the bulk modulus in comparison to the shear modulus. Therefore, as can be seen from Table 2, the hardness of LaRu$_2$As$_2$ is expected to be small and LaRu$_2$As$_2$ should be soft and easily mechinable.

Many important issues related to physical properties of solids can be addressed with Young's modulus. The resistance in opposition to longitudinal tension can be estimated with Young's modulus $Y$. The Young's modulus also manipulates the thermal shock resistance of solids because of having the relation between critical thermal shock coefficient $R$ and $Y$. In fact, the critical thermal shock coefficient $R$ is found to be changed inversely proportional to the Young's modulus $Y$ [48]. The larger the $R$ value, the better the thermal shock resistance. The thermal shock resistance is a fundamental parameter for thermal barrier coating (TBC) materials selection. In comparison to a promising thermal barrier coating material Y$_4$Al$_2$O$_9$ [49], the small $Y$ value of LaRu$_2$As$_2$ signifies that it should have a reasonable resistance to thermal shock.

*3.3 Debye temperature*

Debye temperature $\theta_D$ is a distinctive temperature of crystals at which the highest-frequency mode (and hence all modes) of vibration is excited. It controls many physical properties of materials and is used to make a division between high- and low-temperature regions for a solid. To distinguish the classical and quantum behavior of phonons, the Debye temperature also defines a border line between them. When the temperature $T$ of a solid is elevated over $\theta_D$, every mode of vibrations is expected to be associated with an energy equal to $k_B T$. At $T < \theta_D$, all high-frequency modes seem to be sleep. The vibrational excitations at low temperature are due to acoustic vibrations. The Debye temperature, $\theta_D$, is determined by using one of the standard methods depending on the elastic moduli [50]. This method guides us to calculate the Debye temperature from the average sound velocity $v_m$ by the following equation: $\theta_D = h/k_B [(3n/4\pi) N_A \rho/M]^{1/3} v_m$, where $h$ stands for the Planck's constant, $k_B$ represents the Boltzmann's constant, $n$ is the number of atoms per formula unit, $M$ is symbolized for molar mass,

$N_A$ is the Avogadro's number and $\rho$ denotes the mass density of solid. The average sound velocity within a material is obtained by $v_m = [1/3(1/v_l^3 + 2/v_t^3)]^{-1/3}$. Here $v_l$ and $v_t$ indicate the longitudinal and transverse sound velocities in crystalline solids and are expressed as $v_l = [(3B + 4G)/3\rho]^{1/2}$ and $v_t = [G/\rho]^{1/2}$.

The Debye temperature in cooperation with longitudinal, transverse and average sound velocities calculated within the present formalism is listed in Table 3. The values of $v_l$, $v_t$, $v_m$, and $\theta_D$ found in literature [28] are calculated by using elastic moduli $B$ and $G$ within the Voigt approximation. Here we converted these values into aggregate values using Hill approximations [27]. As a rule of thumb, a lower Debye temperature involves in a lower phonon thermal conductivity. The examined phase $LaRu_2As_2$ has the lower Debye temperature in comparison with other two isostructural compounds as well as a candidate material for thermal barrier coating $Y_4Al_2O_9$ [49] and hence has a lower thermal conductivity. Accordingly, $LaRu_2As_2$ should have opportunity to use as a thermal barrier coating (TBC) material.

**Table 3.** Calculated density ($\rho$ in gm/cm$^3$), longitudinal, transverse and average sound velocities ($v_l$, $v_t$, and $v_m$ in km/s) and Debye temperature ($\theta_D$ in K) of $LaRu_2As_2$.

| Compounds | $\rho$ | $v_l$ | $v_t$ | $v_m$ | $\theta_D$ | Refs. |
|---|---|---|---|---|---|---|
| $LaRu_2As_2$ | 8.799 | 4.731 | 2.676 | 2.976 | 335 | [This] |
| $LaNi_2Ge_2$ | 7.685 | 4.572 | 3.103 | 3.384 | 389 | [28] |
| $LaNi_2P_2$ | 6.853 | 5.210 | 3.684 | 3.994 | 478 | [28] |

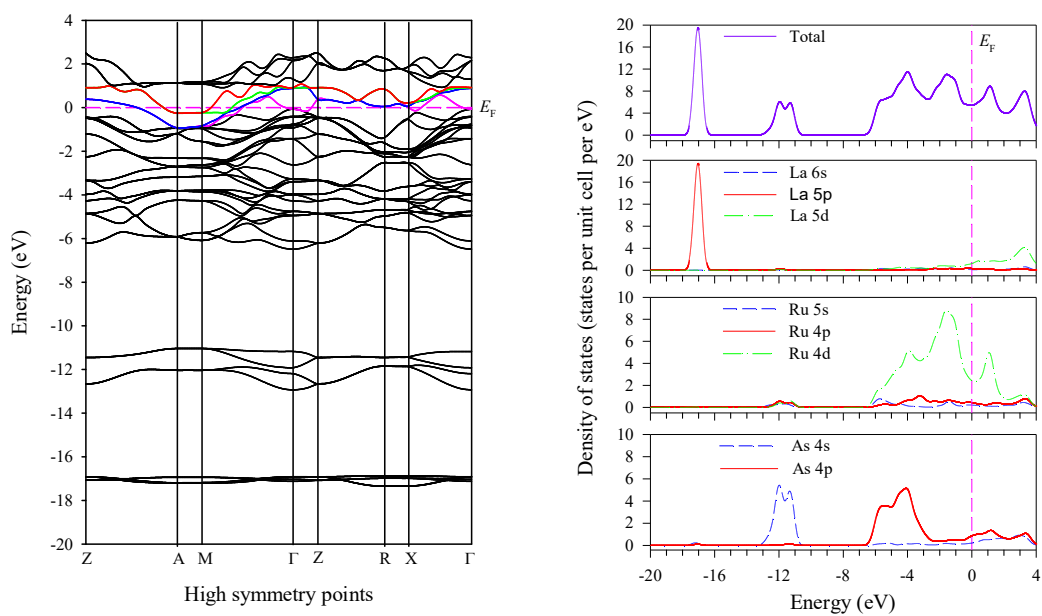

**Fig. 2.** Electronic structures of $LaRu_2As_2$; (a) electronic band structure (left panel), (b) total and partial densities of states (right panel).

*3.4 Band structure and density of states*

Understanding of electronic features (band structure, DOS, etc) is important to explain many physical phenomena, namely optical spectra and transport properties of solids. The full picture of energy bands and band gaps of a solid is known as electronic band structure or simply band structure. Actually, in solid-state and condensed matter physics, the band structure defines certain ranges of energy that are allowed for electrons within a solid, and the ranges of energy that are not allowed for any electrons. The number of bands indicates the available number of atomic orbitals in the unit cell. The calculated energy band structure with high symmetry points in the first Brillouin zone using equilibrium lattice parameters is shown in Fig. 2a. The dashed line in the band structure corresponds to the energy of the highest filled state. This energy level is known as Fermi level, $E_F$. Depending on the position of $E_F$, several important distinctions regarding the expected electrical conductivity of a material can be made. The Fermi level of new 122 phase superconductor takes position just above the valence bands maximum near the Γ point. The new compound should behave metallic in nature since its valence bands cross the Fermi level and overlap noticeably with the conduction bands. In addition, no band gap is found in the Fermi level. The near-Fermi bands show a complicated 'mixed' character, combining the quasi-flat bands with a series of high-dispersive bands intersecting the Fermi level. The detailed features of band structure can be explained with the calculated total and partial densities of states.

The calculated total and partial densities of states are shown in Fig. 2(b), where the vertical broken line denotes the Fermi level, $E_F$. At the left of the Fermi level a deep valley is observed, which is known as pseudogap EP. The delocalized electrons occupy the levels above $E_p$ and the materials containing delocalized electrons turn into metalized. It is seen that the Fermi level of LaRu$_2$As$_2$ lies in a region of relatively low DOS. Consequently, the new compound should be stable in perspective of electronic features as stated in the free electron model. [51]. The new 122 phase superconductor LaRu$_2$As$_2$ should exhibit metallic conductivity due to having finite value of TDOS at $E_F$. The atom resolved partial DOS gives more insights on the electronic structure. Mainly Ru 4d states contribute to the DOS at the Fermi level along with little contribution from La 5d and As 2p states. At $E_F$ the calculated TDOS has a value of 5.4 states/eV-unit cell. The large TDOS is a sign of high metallicity the compound. The TDOS at the Fermi level is quite high compared to some other recently discovered ternary superconducting compounds [52, 53]. This may be responsible for the higher superconducting transition temperature in LaRu$_2$As$_2$, compared to, for example, in ternary silicides [54].

The lowest lying valence bands extended from -17.9 to -16.3 eV arises entirely from La 5p states. These valence bands are observed to be separated by a wide prohibited energy gap of ~3.3 eV form the next valence bands lying in the energy range from -13.0 eV to -10.6 eV. These valence bands are again separated by a forbidden energy gap of width 3.4 eV from the highest valence bands. The highest valence bands contain three visible peaks and broaden in the energy range from -6.6 eV to Fermi level. The left peak originates mainly from As 4p and Ru 4d states. The middle and highest peak are due to equal contribution from Ru 4d and As 4p states. The first peak near to Fermi level forms due to contribution coming from Ru 4d states. In the energy range between -6.6 eV and the Fermi level, the covalent bonding between comprising elements is expected. This is due to the reason that the states are truly degenerate regarding angular momentum and lattice site. Because of the difference in the value of electronegativity among the constituent elements, few ionic interactions can be predicted within LaRu$_2$As$_2$ as well.

*3.5 Mulliken population and Vickers hardness*

The allocation of electrons in several fractional manners among the various parts of bonds can be explained with Mulliken population analysis. The overlap population provides a relation between covalency/ionicity of bonding and bond strength. Sanchez-Portal et al. [55] developed a method for population analysis in CASTEP code using a projection of the plane wave states onto a localized basis. The Mulliken scheme [56] is then applied for population analysis of the resulting projected states. Analyzing the electronic structure calculations with Linear Combination of Atomic Orbital (LCAO) basis sets, this technique is used extensively. Using minimal basis Mulliken scheme [56,57] provides two fundamental quantities regarding atomic bond population are the effective charge and the bond order values between a pair of bonding atoms as follows:

$$Q_\alpha^* = \sum_{i,\alpha} \sum_{n\,occ} \sum_{j,\beta} C_{i\alpha}^{*n} C_{j\beta}^n S_{i\alpha,j\beta}$$

$$\rho_{\alpha\beta} = \sum_{n\,occ} \sum_{i,j} C_{i\alpha}^{*n} C_{j\beta}^n S_{i\alpha,j\beta}$$

where $i, j$ denote the orbital quantum numbers, $n$ indicates the band index, $C_{i\alpha}^{*n}$ as well as $C_{j\beta}^n$ are the eigenvector coefficients of the wave function and $S_{i\alpha,j\beta}$ is the overlap matrix between atoms $\alpha$ and $\beta$.

For understanding the bonding character in crystal systems the effective valence charge and Mulliken atomic population are two important parameters. The effective valence is the difference between the formal ionic charge and Mulliken charge on the anion species within a crystal. It measures the level of covalency/ionicity of chemical bonds. A chemical bond becomes an ideal ionic bond if there is present zero effective valences. In contrast, the positive effective valence indicates the presence of covalent bond and the high values signifies the high level of covalency in chemical bonds. Table 4 lists the effective valence, which reveals the high degree of covalency in chemical bonds inside the new superconductor LaRu$_2$As$_2$. Table 5 tabulates the obtained bond overlap populations between nearest neighbors in this crystal. When the values of overlap population tend to be zero a trivial interaction occurs between the electronic populations of the two atoms. Actually, a weak bond occurs when small Mulliken population exists and such bond makes a negligible contribution to materials hardness. With a low overlap population a bond exhibits high degree of ionicity, whereas with a high value the chemical bond possesses the high level of covalency. Positive and negative bond overlap populations lead to bonding and antibonding states, respectively.

**Table 4.** Population analysis of LaRu2As2.

| Species | Mulliken Atomic populations | | | | | Effective valence Charge (e) |
|---|---|---|---|---|---|---|
| | s | P | d | Total | Charge (e) | |
| As | 1.57 | 3.31 | 0.00 | 4.88 | 0.12 | -- |
| Ru | 2.52 | 6.89 | 7.36 | 16.76 | –0.76 | 3.76 |
| La | 1.92 | 6.01 | 1.77 | 9.71 | 1.29 | 1.71 |

**Table. 5.** Calculated bond and total Vickers hardness $H_v^\mu$, $H_v$ (in GPa) of LaRu$_2$As$_2$ along with bond number $n^\mu$, bond length $d^\mu$ (Å), bond volume $v_b^\mu$ (Å$^3$) and bond and metallic populations $P^\mu$, $P^{\mu'}$.

| Bond | $n^\mu$ | $d^\mu$ | $P^\mu$ | $P^{\mu'}$ | $v_b^\mu$ | $H_v^\mu$ | $H_v$ |
|---|---|---|---|---|---|---|---|
| Ru–Ru | 2 | 2.96437 | 0.98 | 0.11832 | 13.23 | 8.62 | 6.8 |
| Ru–La | 8 | 3.43822 | 1.47 | 0.11832 | 20.64 | 6.44 | |

Hardness assesses the ability of a material to refuse the plastic deformation. The amount of force per unit area acted opposite to plastic deformation takes part in estimating the hardness of a material. F. M. Gao [58] formulated an equation to determine the hardness of non-metallic materials using Mulliken population within first-principles approach. There is no direct relation between delocalized metallic bonding and hardness of materials [59]. For this reason the above method cannot predict the hardness of materials with partial metallic bonding. Gou et al. [60] developed an equation considering a correction for metallic bonding to calculate the bond hardness of mixed metallic crystals as follows:

$$H_v^\mu = 740 \left(P^\mu - P^{\mu'}\right) (v_b^\mu)^{-5/3}$$

where Mulliken overlap population of the $\mu$-type bond is presented by $P^\mu$, $P^{\mu'}$ is a symbol for metallic population and is calculated by using the unit cell volume $V$ and the number of free electrons in a cell $n_{free} = \int_{E_P}^{E_F} N(E)dE$ as $P^{\mu'} = n_{free}/V$, $E_P$ represents the energy at pseudogap, and $v_b^\mu$ stands for the volume of a bond of $\mu$-type, which can be calculated from the bond length $d^\mu$ of type $\mu$ and the number of bonds $N_b^v$ of type v per unit volume through $v_b^\mu = (d^\mu)^3 / \sum_v [(d^\mu)^3 N_b^v]$. For the complex multiband crystals the hardness can be calculated as a geometric average of all bond harnesses as follows [61,62]:

$$H_V = [\prod^\mu (H_v^\mu)^{n^\mu}]^{1/\Sigma n^\mu}$$

where $n^\mu$ represents the number of $\mu$-type bonds, which compose the real multiband crystals. The evaluated Vickers hardness in consideration of positive and reasonable populations between nearest neighbors for the new compound LaRu$_2$As$_2$ is given in Table 5. The hardness value of 6.8 GPa for

LaRu$_2$As$_2$ is very less than that of diamond (96 GPa), which is the hardest material known so far. Therefore, LaRu$_2$As$_2$ is a soft material and it should be easily machineable.

*3.6 Electron charge density and Fermi surface*

To visualize the nature of chemical bonding in LaRu$_2$As$_2$, the electron charge density distribution is calculated and the contour of electron charge density is presented in Fig. 3. In the adjacent scale, the blue and red colors signify the low and high electron densities, respectively. Atom with large electronegativity (electronic charge) exerts a pull on electron density towards itself [63]. Because of large difference in electronegativity and radius of atoms, the electronic charges around Ru (2.20) and As (2.18) are greater than that around La (1.10). The higher electronegativity of Ru and As exhibits strong accumulation of electronic charge while relatively low charge density indicates weak charge accumulation of La. Typical ionic picture based on standard oxidation states of atoms can help to explain the bonding nature of compounds. In LaRu$_2$As$_2$, the oxidation states of atoms are La$^{2+}$, Ru$^{2+}$ and As$^{3-}$. The sum of the oxidation numbers must be zero for a neutral compound such as LaRu$_2$As$_2$. In LaRu$_2$As$_2$ structure, the [Ru$_2$As$_2$] blocks are separated by La atomic sheets. The charge states of these sheets/blocks are as follows: [La]$^{2+}$ and [(Ru$^{2+}$)$_2$(As$^{3-}$)2]$^{2-}$. It can be said that a charge transfer will take place from [La]$^{2+}$ sheets to [Ru$_2$As$_2$]$^{2-}$ blocks. The existence of ionic bonds between Ru-As within [Ru$_2$As$_2$]$^{2-}$ blocks can be guessed from negative and positive charge balance at the positions of relevant atoms. From atom resolved density of states it is observed that the hybridization between Ru 4d and As 4p states lead to formation of covalent Ru-As bonds. Moreover, metallic type Ru-Ru bonds are expected to exist in [Ru$_2$As$_2$] blocks due to overlapping of Ru 4d states near Fermi level [Fig. 2b]. Surprisingly, no As-As bonds exist between two adjacent [Ru$_2$As$_2$] blocks in LaRu$_2$As$_2$. But such bonds are found to be present in (Ca/Ba)Fe$_2$As$_2$ iron pnictides [64]. The possible reason for this contravention may be explained as: the low strength of Fe-As bonds in iron pnictides in comparison to Ru-As bonds in LaRu$_2$As$_2$ favors to form As-As bonds in (Ca/Ba)Fe$_2$As$_2$. The above discussion reflects that the chemical bonding in LaRu$_2$As$_2$ can be described as a highly anisotropic combination of ionic, covalent and metallic interactions.

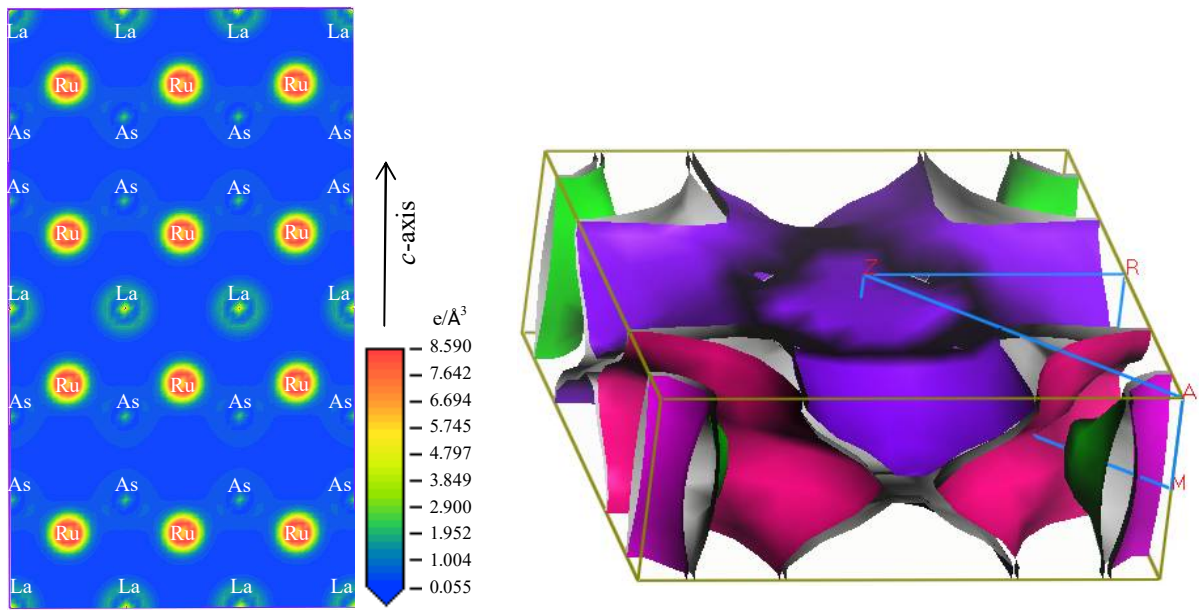

**Fig. 3.** Charge density (left) and Fermi surface (right) of LaRu$_2$As$_2$.

The Fermi surface topology has been calculated, which is presented in Fig. 3 (right). As we observe, the near-Fermi surface band pictures have a complicated 'mixed' character: simultaneously with quasi-flat bands and high-dispersive bands intersects the Fermi level. These features concede a multi-sheet Fermi surface consisting of two quasi-two-dimensional electron-like sheets in the corners of the Brillouin zone. The nearest sheet is also parallel to A-M direction and the distant sheet is half-fold and intersected at the centre of the sheet by the Γ-M line. Between these two sheets a concave sheet appears and the Γ-M line serves as the axis of this tree-dimensional sheet. The central electron-like sheet is very complicated and has four wings shaped of open half-tube cutting along own axis. These four wings form a red-cross shape keeping each arm along Z-R direction. The existence of the multiband nature suggests that the new compound $LaRu_2As_2$ could be a class of multiple-gap superconductor.

*3.7 Electron-phonon coupling constant*

Accurate value of electron-phonon coupling constant (λ) is important to evaluate the superconducting transition temperature $T_c$. The method assigned to QUANTUM ESPRESSO program [65] can accurately compute the λ value directly. We stress that a double-delta function integration over a dense net of electron and phonon vectors ($k$ and $q$) is required to compute the electron-phonon coupling constant with significant computational resources. Moreover, this method is more complicated for a crystal system containing many atoms in its unit cell like $LaRu_2As_2$. All of these prohibited us to carry on with this method. We have an alternative method in which the equation $1 + \lambda = 3\gamma/[2\pi^2 k_B^2 N(E_F)]$ provides the electron-phonon coupling constant [66]. But in the present situation we have no experimental values for electronic specific-heat coefficient γ and theoretically calculated value yields a lower value of λ [67]. So, we have only one indirect method using the McMillan's equation:

$$\lambda = \frac{1.04 + \mu^* \ln\left(\frac{\theta_D}{1.45 T_c}\right)}{(1 - 0.62\mu^*)\ln\left(\frac{\theta_D}{1.45 T_c}\right) - 1.04}$$

The repulsive Coulomb potential $\mu^*$ is assigned a value in the range of 0.10-0.15 and the choice of $\mu^*$ is fairly arbitrary. Using known $T_c$ = 7.8 K [11] with our calculated Debye temperature $\theta_D$ = 335 K in McMillan's equation, we have λ = 0.64 and 0.76 for $\mu^*$ = 0.10 and 0.15, respectively, which implies that $LaRu_2As_2$ should be a moderately coupled BCS superconductor.

**4. Conclusions**

DFT methodologies were used within the first-principles pseudopotential method to calculate some physical properties of newly discovered superconductor $LaRu_2As_2$. The evaluated lattice constants show sound agreement with the experimental results. The obtained single crystal elastic constants obey the mechanical stability conditions for the new tetragonal system. The $LaRu_2As_2$ crystal shows high uniaxial elastic anisotropy. The newly synthesized compound is expected to resist the shear distortion in the <110> direction than in the (100) plane. Cauchy pressure, Poisson's ratio as well as Pugh's ratio indicate that $LaRu_2As_2$ should be a ductile material. The relatively small young's modulus indicates that $LaRu_2As_2$ should attain reasonable resistance to thermal shock. The low Debye temperature also suggests that this compound should be favorable for use as a thermal barrier coating material. The calculated electronic structures reveal that the intra-atomic bonding in new 122 phase compound $LaRu_2As_2$ may be explained as a mixture of ionic, covalent and metallic interactions. The hardness value of 6.8 GPa implies that this material should be soft and easily machineable. The obtained Fermi surface exhibits the multi-band nature, suggesting that the new ternary compound $LaRu_2As_2$ could be a kind of multiple-gap superconductor. The calculated λ indicates the strong electron-phonon coupling in $LaRu_2As_2$.

Very recently, it came to our notice that Rahaman and Rahman has done some *ab initio* calculations on $LaRu_2As_2$ and $LaRu_2P_2$ compounds [68]. Some of the basic features of those calculations agree quite well with the ones presented in this paper.

To conclude, we expect that this study will stimulate further experimental and theoretical research activities on the newly synthesized $LaRu_2As_2$ compound.